\documentclass[11pt]{article}

\usepackage{amsfonts, amsmath}
\usepackage[margin=1in]{geometry}
\usepackage{natbib}
\title{At what time does a quantum experiment have a result?}

\author{Thomas Pashby\footnote{Department of Philosophy, University of Chicago, 1115 E. 58th St. 
Chicago, IL 60637. pashby@uchicago.edu}}

\begin{document}

\maketitle

\section{Introduction}

This paper provides a critical guide to the literature concerning the answer to the question: when does a quantum experiment have an result?  This question was posed and answered by Rovelli (\cite{rovelli1998incerto}) and his proposal was critiqued by Oppenheim, Reznick and Unruh (\cite{oppenheim2000does}), who also suggest another approach that (as they point out) leads to the quantum Zeno effect. What these two approaches have in common is the idea that a question about the time at which an event occurs can be answered through the instantaneous measurement of a projector (in Rovelli's case, a single measurement; in that of \cite{oppenheim2000does}, a repeated measurement).  However, the interpretation of a projection as an instantaneous operation that can be performed on a system at a time of the experimenter's choosing is problematic, particularly when it is the time of the outcome of the experiment that is at issue.

In classical probability theory, illustrated in Section \ref{ex_exp} with the simple case of an exponential decay law, if the time of an event is a random variable (to be determined by the result of experiment) then one integrates over a probability density to find the  probability for the occurrence of the event within a time interval (not at an instant).   This suggests treating a time-indexed (Heisenberg picture) family of projectors as supplying a probability density for occurrence at $t$, rather than probabilities for the result of a measurement at $t$.  Treating the time of occurrence as a variable whose value is to be determined by experiment, Section \ref{cond_prob} provide a general method for defining a generalized quantum observable (or POVM) that supplies properly normalized \emph{conditional} probabilities for the time of occurrence.  This method resolves the difficulties of the previous approaches, discussed in detail in Section \ref{comparison}.

\section{Do we choose the `time of collapse'?}\label{time_collapse}

The previous attempts to address the question posed here made use of the idea that a quantum measurement is an instantaneous process that changes the state of the system.  It is habitually assumed in discussions of quantum foundations that the observer is free to choose which observable to measure.\footnote{Some surprisingly powerful results have followed from this assumption. Consider the role of parameter independence in Bell's Theorem or, more controversially, the so-called Free Will Theorem \citep{conway2006free}.}  However, the freedom of choice allowed by the quantum formalism concerning the \emph{time} at which the observable is measured is seldom emphasized.  It is this second kind of freedom that I aim to challenge.

In orthodox quantum mechanics a self-adjoint operator $A$ corresponds to a physical quantity that can be measured by an experiment, i.e., a Schr\"odinger picture observable.  In the Heisenberg picture, the system state remains constant while the observables vary with time, which means that to measure $A$ at time $t$ we must choose $A_t= U^\dag_t A U_t$  from a whole family of time-indexed observables (each self-adjoint), each of which measures $A$.   This choice matters:  in general, $A_t$ and $A_{t'}$ will have different expectation values for the same state (when $t\neq t'$).  But since nothing in the formalism determines at what time the observable is to be measured, the experimenter is apparently able to choose freely at what time she will act on the system by measuring $A_t$, for some $t$.

This idea can be traced back to the first edition of Dirac's classic \emph{Principles of Quantum Mechanics}.  Working in the Heisenberg picture, Dirac describes the situation as follows:
\begin{quote} 
A system, when once prepared in a given state, remains in that state so long as it remains undisturbed. \dots [I]t is sometimes purely a matter of convenience whether we are to regard a system as being disturbed by a certain outside influence, so that its state gets changed, or whether we are to regard the outside influence as forming part of and coming in the definition of the system \dots There are, however, two cases where we are in general obliged to consider the disturbance as causing a change in the state of the system, namely, when the disturbance is an observation and when it consists in preparing the system so as to be in a given state. \dots This requires that the specification of an observation shall include a definite time at which the observation is to be made \dots \cite[p.~9]{dirac1930principles}
\end{quote}

There is, however, a real disconnect here between experimental physics and the quantum formalism.  In a typical quantum experiment one uses a detector (or several) to collect results, and in practice an experimenter has no real control over the time at which the detector will detect something (by clicking, or fluorescing, or what have you).  In a classic two-slit experiment, for example, the screen lights up in a fairly well-defined place at a fairly-well defined time and there is little that the experimenter can do to influence the location of this detection event in time (or space).

Now, the time at which a quantum particle (e.g., a photon) is \emph{emitted}  can be fairly tightly controlled (by a pulsed laser) and this control allows for measurement of time of arrival, often used as a proxy for energy.  But the time of arrival (i.e., the time of detection minus the time of emission) is not the sort of thing that the experimenter can control with precision, and the kind of control that one might exert (by increasing the energy, say) would be best described theoretically as changing the state.   In practice, it seems there is nothing the experimenter may choose to do in order to `make a measurement' at a specific instant of time;  there is no \emph{action} that the experimenter performs to bring about the detection of a particle---these detection events happen of their own accord.

\section{Which observables could measure times?}\label{what_instead}

The discussion above indicates that the orthodox interpretation of the formalism is on the wrong track. At the root of the problems with trying to make sense of the `time of measurement' or `time of collapse' in quantum theory is, I contend, the idea that the experimenter has control of precisely \emph{when} to apply an observable.   My suggestion is that we can avoid these problems by asking questions instead about the time of detection (or registration)---a time that actually can be recorded in a quantum experiment.  This time may be taken as a proxy for the time of a microscopic event (like the ionization of an atom, say) but the key point is that predictions made for the distribution in time of these events can be compared with actual experimental statistics.

To correctly describe these experimental statistics requires a probability distribution for the time of detection.  In classical probability theory, this means treating the time of detection as a random variable.  In quantum theory, then, one must treat the time of detection as an observable.  Now, as is well-known, Pauli's Theorem implies that there is no self-adjoint operator with the requisite properties (\cite{srinivas1981time}).  But what is actually established by this result is that if the Hamiltonian has a semi-bounded spectrum then there can exist no time-translation covariant Projection Valued Measure (PVM).  However, (as is also well-known) this does not prevent the use of generalized observables, known as Positive Operator Valued Measures (POVMs) to represent the time of an event (\cite{busch1994time,egusquiza2002standard}).\footnote{In particular, requiring that an observable must correspond to a self-adjoint operator (and thus a PVM) is sufficient to guarantee that it returns a valid probability distribution, but this requirement is not necessary.  What \emph{is} necessary for an observable to return a valid probability distribution from the quantum state is that it defines a POVM (\cite{busch2003quantum}), and every PVM is a POVM. Given Pauli's Theorem, then, an event time observable will have to be a POVM that is not a PVM (and thus its first moment defines an operator that is symmetric but not self-adjoint).} The problem one then faces is to pick out a particular POVM as being empirically significant.

Spatiotemporal symmetries have traditionally played a significant role in attempts to pick out experimentally relevant POVMs. \cite{wightman1962} argued that the position observable could be (in effect) uniquely determined by requiring that the projectors of the associated PVM provide a representation of the Euclidean group. Over twenty years later, \cite{Werner1986} showed how to generalize this method to find POVMs appropriate to the symmetry group of any given three-dimensional hyperplane (spatial or spatiotemporal).  In doing so, he subsumed both the position observable and the quantum time of arrival (see \cite{kijowski1974,egusquiza2002standard}) within the more general notion of a screen observable.

Inspired by \cite{mackey1963mathematical}, Wightman had interpreted the projectors of his position PVM as `experimental questions'.  Underlying this idea is the notion that a projector, $P$, represents a (possible) property of the system. An experimental question asks: ``Does the system have property $P$?''  Each projector $P$ defines a Heisenberg picture family of observables $P_t= U_t^\dag P U_t$ corresponding to asking this experimental question of the system at time $t$. Here, again, the time at which the question is asked is apparently the free choice of the experimenter. 

In the case of the position PVM, $\Delta \mapsto P_\Delta$, a projector $P_\Delta$ corresponds to the property of being located (or ``localized'') in spatial region $\Delta$. According to Wightman, measuring the observable $P_\Delta$ at time $t$ asks the experimental question, ``is the system located in $\Delta$ at time $t$?''  However, from the perspective of an actual experiment---operationally, that is---this interpretation is problematic.  If we interpret $P_\Delta$ in terms of a detector located in the region $\Delta$, this question asks: ``if the detector, sensitive in region $\Delta$, is turned on for an instant $t$, does it register the system?''   But very few experiments (if any) involve detectors that are sensitive for a mere instant of time.

Much more common are detectors that are sensitive over an \emph{extended} period of time, like a luminescent screen.  With Werner's definition of a screen observable, this is modeled by a spatiotemporal three-dimensional hyperplane that has one dimension of time and two of space\footnote{That is, Werner models a screen as a two-dimensional spatial plane, $\Sigma$, that extends indefinitely in time. } which leads to a POVM, $\Sigma\times I \mapsto E_{\Sigma\times I}$. Following  Wightman, measurement of $E_{\Sigma\times I}$ could be thought of as asking, ``if the detector, sensitive in area $\Sigma$, is turned on for an interval of time $I=[t_1,t_2]$, does it register the system?'' 

However, applying this interpretation in the context of an actual experiment presents a puzzle.   Under this interpretation, the expectation value $\langle  E_{\Sigma \times I} \rangle_\psi$ gives the probability for registration during $I$ in state $\psi$, given that the detector is turned on for interval $I$.  Therefore, the probability of detection during $I$ and the probability of no detection during $I$ must sum to one. Since the POVM $\Sigma \times I \mapsto E_{\Sigma\times I}$ is properly normalized, it is indeed the case that $E_{\Sigma\times I} + E_{\Sigma\times (\mathbb{R} - I)} =\mathbb{I}$.  But $E_{\Sigma\times (\mathbb{R} - I)}$ corresponds to detection in region $\Sigma$ at some time $t$ not in $I$, and the detector can hardly be expected to detect the system when it is switched off.

This shows that the condition under which these probabilities apply is \emph{not} the choice to turn on the detector for time interval $I$.  A better interpretation is given in terms of a screen detector that is always on, in which case $\langle  E_{\Sigma \times I} \rangle_\psi$ gives the probability of detection during $I$ rather than $\mathbb{R} - I$.   Thus, because of the normalization condition $E_{\Sigma\times \mathbb{R}} = \mathbb{I}$, this POVM describes an experiment in which the detector is left on until it detects something.  Another way of saying this: the probabilities $\langle  E_{\Sigma \times I} \rangle_\psi$ are \emph{conditional} probabilities, conditioned on the fact that the detector, sensitive in the plane $\Sigma$, does fire at some time $t\in \mathbb{R}$.

Some years later, \cite{brunetti2002time} provided the means to generalize the definition of screen observables to (what they call) time of occurrence observables. Instead of interpreting $P_\Delta$ as a property of the system (the property of being located at region $\Delta$) and $P_\Delta(t)$ as an experimental question asked at time $t$,  they suggest interpreting $P_\Delta$ (an effect) as an \emph{event} (here, the event of the detector firing in region $\Delta$) so that $P_\Delta(t)$ represents the occurrence of that event at time $t$.  The normalization of the time of occurrence POVM that results, $\Delta\times I \mapsto E_{\Delta\times I}$, again indicates that this observable describes a detector that is always on.\footnote{Note that this can be seen to subsume the idea of a screen observable: \cite{hoge2008} explicitly demonstrates that this returns the appropriate screen observable as the volume $\Delta$ becomes an area, $|\Delta|\rightarrow 0$.}  

Here, the probabilities are conditioned on the fact that the detector, sensitive to spatial region $\Delta$, does fire at some time $t\in \mathbb{R}$; this is what the normalization condition $E_{\Delta\times \mathbb{R}} = \mathbb{I}$ represents. This interpretation of the time of occurrence shows how to avoid the problematic assumption that the time of measurement is under the control of the experimenter: by normalizing using the condition that the experiment has an outcome at \emph{some} time, we obtain a distribution of conditional probabilities for the time at which that outcome occurs (i.e., probabilities that apply given that the detector is left on until it detects something).  

\subsection{Conditional Probabilities, POVMs and L\"uders' Rule}\label{luders_problem}

Conventionally, conditional probabilities arise in quantum mechanics by conditioning the state using L\"uders' Rule.  In terms of the density operator for a system, $\rho$, L\"uders' Rule states that the conditional probability of a positive result for a projector $E$ given a positive result for a projector $F$ is
\begin{equation}\label{luders_rule}
	\Pr ( E | F ) = Tr [ E \rho' ] = \frac{ Tr [ F E F \rho ] }{Tr [ F \rho ] },
\end{equation}
and thus the following state transition is associated with this conditionalization:
\begin{equation}
\rho \rightarrow \rho'  =  \frac{ F \rho F }{Tr [ F \rho ] }.
\end{equation}

However, the time of occurrence necessarily defines a POVM rather than a PVM (due to Pauli's Theorem) so it cannot supply the projectors to which L\"uders' Rule would apply.  When dealing with such a POVM, the standard move is to make use of the associated L\"uders Operation instead, which conditions the system using a positive operator (\cite{busch1995operational}). The L\"uders Operation is obtained simply by replacing the projector $E$ with the positive operator\footnote{Any bounded positive operator $A$ has a unique positive square root $A^{1/2}$ such that $\left(A^{1/2}\right)^2 = A$. A projector $P$ is just a positive operator for which $P^2 = P$ and thus $P^{1/2} = P$.} $A^{1/2}$:
\begin{equation}\label{luders_op}
 \rho \rightarrow \hat{\rho} = \frac{ A^{1/2} \rho A^{1/2} } {Tr [ A \rho ]  }.
\end{equation}

Naively, one might expect the following expression to give a conditional probability for $B$ given $A$ (both effects)
\begin{equation}
W( B | A ) = Tr [ B \hat{\rho} ] =  \frac{ Tr [A^{1/2} B A^{1/2} \rho ]} {Tr [ A \rho ]  }.
\end{equation}
However, for $W(B | A)$ to be a well-formed conditional probability, it must be the case that $W (A | A) =1$, i.e., that the probability of $A$ given $A$ is one. But since
\begin{equation}
W( A | A ) = Tr [ A \hat{\rho} ] =  \frac{ Tr[A^{1/2} A A^{1/2} \rho] } {Tr [ A \rho ]  } =  \frac{ Tr [A^2   \rho ]} {Tr [ A \rho ]  }
\end{equation}
this only obtains if $A^2 = A$, in which case $A$ is a projector and the L\"uders Operation reduces to L\"uders' Rule.

Therefore, if the L\"uders Operation is to lead to a well-formed conditional probability for the time of occurrence POVM, it must do so by other means.\footnote{I first raised this difficulty in \cite{pashby2015time}, which also contains a history of time in early quantum theory.} I resolve this problem in Section \ref{cond_prob} by showing how to derive a valid conditional probability distribution from the Born Rule (applied at an instant) that supplies an alternative way to relate the time of occurrence POVMs of Brunetti and Fredenhagen to the L\"uders Operation.  This approach can be motivated by considering how probabilities for the time of occurrence arise in the following simple case.

\section{An Example: Exponential Decay}\label{ex_exp}

Here we examine how classical probability theory provides the probability distribution for the time of an event whose occurrence follows an exponential decay law.  For a radioactive atom, one might justify such a law phenomenologically  as follows.  Taking a uniform sample of a radioactive isotope we observe that the number of nuclei that will decay in a given interval of time is proportional to the original number of nuclei in the sample.  Furthermore, the proportion of the original sample of $N_0$ nuclei that remains after a time $t$ is seen to follow the simple rule
\[   \frac{N_t}{N_0} = e^{-t/T} = e^{-\lambda t}, \]
where $T$ is the mean lifetime of the parent nuclei.

The half-life, $\tau_{1/2}$, is the time it takes for half the sample to decay, i.e., the value of $t$ such that $e^{-\lambda t}$ is equal to a half.  The half-life of radium, for example, is approximately 1600 years.  The expected proportion of the sample that remains at time $t$, then, can be equivalently expressed as
\[ \frac{N_t}{N_0} = e^{- (\tau_{1/2}t)/ \ln 2} = e^{-\lambda t}.\]
And if our sample consists of a \emph{single} atom then this fraction may be interpreted as the probability of decay after time $t$.

To calculate the probability that the time of decay of a single atom, $t_d$, lies within some time interval $[t_1,t_2]$, where $t_1\geq0$ and $t_2>t_1$, we integrate the corresponding \emph{probability density}, $\lambda e^{-\lambda t}$, as follows:
\begin{equation}\label{exp_prob}
 \Pr(t_1\leq t_d < t_2) = \int_{t_1}^{t_2} \lambda e^{-\lambda t} dt = \frac{1}{T}  \int_{t_1}^{t_2} e^{-\lambda t} dt.
 \end{equation}

Operationally, these probabilities refer to the result of an experiment in which an atom is held within a detector until the detector registers the emission of a decay product at time $t_d$.  If we imagine the data from an ensemble of decay experiments, each with an identical atom, then we would expect to see a distribution in time for the corresponding variable $T_d$ (of which $t_d$ is a value) that approximates the exponential decay law.  To collect the data for this ensemble, each run of the experiment must continue until a detection event is observed, and the outcome of each such experiment is \emph{a time}---the actual time of decay, $t_d$.  These times $t_d\in\mathbb{R}$ are the values of a random variable $T_d$ (in the sense of probability theory) corresponding to the time of decay.

Using this assignment of probabilities to time intervals, we have the means to answer the question: ``Has the atom already decayed?''  To this question, asked at time $t$, there corresponds the probability assignment
\[  \Pr (0 \leq t_d < t ) = \int_0^t \lambda e^{-\lambda t} dt ,\]
in which case we also have a straightforward interpretation of the normalization condition:
\begin{equation}\label{exp_prob}
 \Pr(0\leq t_d < \infty) = \int_0^\infty \lambda e^{-\lambda t} dt = 1.
 \end{equation}
This simply says that the decay is certain to occur during the time interval $[0,\infty)$. 

Note, therefore, that it is simply built into this assignment of probabilities \emph{that the atom will decay}, which shows that these are \emph{conditional} probabilities, as I now explain.  Consider the following simple example.  In assigning probabilities to the results of a coin toss occurring at some later time $t$, the obvious normalization implies that the probability of the event that \emph{either} the coin lands heads \emph{or} the coin lands tails is 1, and thus the coin toss is certain to occur.\footnote{This is just the assumption that we have a well-formed event space.}  Therefore, if this probability assignment is correct, then the future occurrence of the coin toss is guaranteed.

This may make it sound as if we have ensured the future occurrence of this event simply by assigning probabilities to its results, but of course we have not. All this says is that the probabilities are \emph{conditioned} on the occurrence of the coin toss and apply just in case it does take place, though it need not---if the coin toss does not occur then the condition is not met and there are no events to which the probability assignment applies.

In the case of the time of decay the only difference is that the events (in the sense of probability theory) to which we assign probabilities describe the decay of the atom at different time intervals.  These are the possible results of the decay process, and to the event of decay at \emph{some} time we assign probability one.   But, again, this is just to say that this is an assignment of \emph{conditional} probabilities.  No event is forced to happen simply by making a probability assignment conditioned on its occurrence, and if in fact no decay subsequently occurred then there were simply no events to which the conditional probability assignment applied.   

\section{The Time of an Outcome as a Conditional Probability}\label{cond_prob}

Taking the above treatment of the time of decay as a model, I now show how to arrive at a well-formed probability distribution for the time of occurrence by treating the real number supplied by the Born Rule as a probability density.   We begin with a time-indexed family of Heisenberg projectors $P_t$, where $P$ represents the occurrence of the outcome in question (such as the click of a detector).  These projectors correspond to a family of propositions stating that ``the outcome occurs at $t$''.  We will assume that an outcome occurs at some $t\in\mathbb{R}$, and only once.\footnote{It must be admitted that this restricts the application of this technique to experiments where only a single outcome of some type is expected.  However, these experiments are precisely those where probabilistic answers to the question ``when does the experiment have an outcome?'' makes sense.}  With this assumption the occurrence of an outcome at $t$ and the occurrence of an outcome at $t'$ are mutually exclusive events (for $t\neq t'$).

These assumptions amount to the claim that we are dealing with conditional probabilities, conditioned on the fact that the experiment has a unique outcome at some unknown time.  This condition will be reflected in the proper normalization of our probability distribution by means of a suitably formed probability density.  Assuming the Heisenberg state of the system is a pure state $|\psi\rangle$, an application of the Born Rule to the Heisenberg projectors $P_t$ returns the following function of $t$:
\begin{equation}
	f(t) = \langle \psi | P_t \psi \rangle . 
\end{equation}
This is not yet suitable for use as a probability density; if we integrate $f(t)$ over $t$ we find
\begin{equation}
	\int_{-\infty}^\infty f(t) dt = \int_{-\infty}^\infty \langle \psi | P_t \psi \rangle dt, 
\end{equation}
which is evidently unnormalized and may not even converge.  However, if the integral is finite (at least for some $|\psi\rangle$) we may use it as a normalization factor in defining the genuine probability density
\begin{equation}
	f'(t) = \left(\int_{-\infty}^\infty f(t) dt\right)^{-1} f(t) = \mu f(t),
\end{equation}
where $\mu$ now resembles the normalization factor $\lambda  = 1/T$ of an exponential decay (where $T$ is the mean decay time).

In quantum mechanics, one would expect the mean time of occurrence to be provided by the expectation value of some operator.  The linear operator $S$ defined by varying $\phi , \psi \in \mathcal{H}$ in the following expression immediately suggests itself for this purpose:\footnote{Note that this operator depends on the Hamiltonian $H$ through the Heisenberg picture family $P_t= U_t^\dag P U_t$, where $U_t=e^{iHt}$. It is not surprising that an operator related to the time of an event should depend on the dynamics.}
\begin{equation}
	\langle \phi | S \psi \rangle := \lim_{\tau\rightarrow \infty} \int_{-\tau}^\tau \langle \phi | P_t \psi \rangle dt .
\end{equation}

Where defined, this operator $S$ is positive and thus symmetric (on an appropriate domain).   Note that there is a good reason to be interested only in those states for which the expectation of $S$ is strictly positive and finite (otherwise we are dealing with an experiment without an outcome), in which case we may write:\footnote{For reasonable choices of $P$ and $H$,  \cite{brunetti2002time} demonstrate that this definition does lead to a positive operator $S$ defined on the domain obtained by taking the orthogonal complement of the subspace of states for which the expectation of $S$ is $0$ or infinity.}
\begin{equation}
\mu = \frac{ 1} {\langle S \rangle_\psi } = \frac{ 1 } { \langle \psi | S \psi \rangle }.
\end{equation}

This allows for the definition of a probability distribution for the time of the outcome, $t_o$, which depends on the state $|\psi\rangle$.  If $f'(t)$ is the correct probability density then in the state $|\psi\rangle$ the probability that the statement ``the outcome occurs during $[t_1,t_2]$'' is true is given by
\begin{equation}
 \Pr(t_1 \leq t_o < t_2) = \int_{t_1}^{t_2} f'(t) dt = \frac{1}{\langle S \rangle_\psi } \int_{t_1}^{t_2} \langle \psi | P_t \psi \rangle dt.
\end{equation}
Again, in quantum mechanics one would expect that this probability is the expectation of a positive operator, and the linear operator $F_{[t_1,t_2]}$ defined by varying $\phi , \psi \in \mathcal{H}$ in the following expression suggests itself for this purpose:
\begin{equation}
 \langle \phi | F_{[t_1,t_2]} \psi \rangle := \int_{t_1}^{t_2} \langle \phi | P_t \psi \rangle dt.
\end{equation}
This operator is positive and bounded, and one can see that the operators $F_{[t_1,t_2]}$ obtained by varying $[t_1,t_2]$ are apt to form an unnormalized POVM, $\Delta \mapsto F_\Delta$, of which $S$ is the first moment, i.e.\ $F_\mathbb{R} = S$. Using this operator, we can rewrite the probability for time of occurrence as 
\begin{equation}\label{time_cond_prob}
 \Pr(t_1 \leq t_o < t_2) = \int_{t_1}^{t_2} f'(t) dt = \frac{\langle \psi | F_{[t_1,t_2]} \psi \rangle }{\langle \psi | F_\mathbb{R} \psi \rangle } .
 \end{equation}
 
In quantum mechanics, however, probabilities are given by the expectations of a normalized POVM $\Delta \mapsto E_\Delta$, for which $E_\Delta \leq \mathbb{I}$ and $E_\mathbb{R} = \mathbb{I}$.  To see how such a POVM arises naturally from the expression above, we first consider how $S$ acts to condition the state through its associated L\"uders Operation (Eqn.\ \ref{luders_op}),
\begin{equation}
 \rho \rightarrow \hat{\rho} = \frac{ S^{1/2} \rho S^{1/2} } {Tr [ S \rho ]  }.
\end{equation}
In Section \ref{luders_problem}, I pointed out that a naive interpretation of the L\"uders Operation does not lead to a conditional probability.  We have avoided this problem here by adopting the expression above (\ref{time_cond_prob}) as our definition of a conditional probability.  Indeed, applied to the state $\rho$, (\ref{time_cond_prob}) leads to:
\[ \Pr (t_1 \leq t_o <t_2 ) = W' ( F_{[t_1,t_2]} | S ) := \frac{ Tr [ F_{[t_1,t_2]} \rho ] } {Tr [  S \rho ] }. \]

Here the probabilities are properly conditioned since obviously $W' ( F_\mathbb{R} | F_\mathbb{R} ) = 1$, although the connection to L\"uders' Rule is not yet clear.  This connection is provided by making use of Brunetti and Fredenhagen's operator normalization technique, which converts $\Delta \mapsto F_\Delta$ to a properly normalized POVM $\Delta \mapsto E_\Delta$ by defining
\[ E_{[t_1,t_2]} := S^{-1/2} F_{[t_1,t_2]} S^{-1/2} ,\]
where $S^{-1/2}$ is the inverse of $S^{1/2}$.

We now obtain an equivalent expression for our conditional probability by applying the L\"uders Operation for $S$ to the state, which corresponds to detection at some time, and then taking the expectation of $E_{[t_1,t_2]}$.  That is, the expression
\begin{equation}
 \Pr (t_1 \leq t_o < t_2 ) = \Pr ( E_{[t_1,t_2]} | S) = Tr [E_{[t_1,t_2]} \hat{\rho} ]  = \frac{ Tr [ S^{1/2} E_{[t_1,t_2]} S^{1/2} \rho ] } {Tr [ S \rho ] } = W' ( F_{[t_1,t_2]} | S ),
 \end{equation}
 returns the probability for the occurrence of the outcome described by the projector $P$ during the time interval $[t_1,t_2]$.  This probability is conditioned on the occurrence of the outcome at \emph{some} time $t_o\in \mathbb{R}$ during the entire course of the experiment.\footnote{In case this seems only tangentially related to L\"uders' Rule, note that in the temporally extended Hilbert space this expression becomes identical to L\"uders' Rule. See \cite{pashby2014phd}, \S8.3.}
 
\section{Comparison with Previous Approaches}\label{comparison}

In Section \ref{time_collapse}, I argued that the orthodox account of quantum mechanics leads to the idea that the measurement of an observable is an instantaneous process that takes place at the behest of the experimenter.  By rejecting this idea in favor of an interpretation of time-indexed Heisenberg projectors as describing the occurrence of a detection event we were led to define a valid probability distribution (conditioned on the occurrence of that event during the course of an experiment).  This removed the problematic assumption that the observer may bring about an outcome at the time of her choice.  Rovelli's proposal, although ingenious, amounts to simply embracing that idea.

\subsection{Rovelli's `One Shot' Approach}

\cite{rovelli1998incerto} defines a projector $M$ whose measurement at time $t$, he says, provides an answer to the question: ``Has the measurement happened?''  This projector is defined on the (tensor product) Hilbert space of system and apparatus and has as its range the product states indicating a perfect measurement, i.e., those joint states displaying a complete correlation of microscopic states with the corresponding `pointer' states of the macroscopic apparatus. 

Following Rovelli's presentation, we consider a simple measurement of a discrete observable with two eigenvalues $\{a,b\}$.  To each eigenvalue there corresponds a subspace of the state space of the object system and the measurement interaction correlates these states with states of the apparatus that indicate those values.  If the initial state of the combined system is
\begin{equation}
\Psi(0) = (c_a |a\rangle + c_b |b\rangle) \otimes |init\rangle,
\end{equation}
where $|c_a|^2 + |c_b|^2=1$ and $|init\rangle$ is the `ready' state of the apparatus, then the final, post-measurement state is
\begin{equation}
 \Psi(T) = c_a |a\rangle \otimes |O_a\rangle + c_b | b \rangle \otimes |O_b\rangle,
\end{equation}
where $|O_a\rangle$ and $|O_b \rangle$ are states of the apparatus that indicate results $a$ and $b$, respectively.

Rovelli defines his projector $M$ by specifying its action as follows:
\begin{eqnarray}
	M(|a\rangle \otimes |O_a\rangle) = |a\rangle \otimes |O_a\rangle \\
	M(|b\rangle \otimes |O_b\rangle ) = |b\rangle \otimes |O_b\rangle \\
	M (|a\rangle \otimes |O_b\rangle )= 0 \\
	M ( |b\rangle \otimes |O_a\rangle ) = 0
\end{eqnarray}
and if $|\phi\rangle$ is a state of the apparatus such that $\langle \phi | O_a \rangle =0$ and $\langle \phi | O_b \rangle=0$ then
\begin{equation}
	M(|\psi \rangle \otimes |\phi\rangle) =0
\end{equation}
for any state of the system, $|\psi\rangle$.

Applying the usual interpretation to the measurement of a projector (a self-adjoint operator with eigenvalues $\{0,1\}$) there are two possible results: either the states of system and apparatus are perfectly correlated (eigenvalue 1), or they are not (eigenvalue 0).  Rovelli interprets these results as answering the question ``has the measurement happened?'', to which a measurement of $M$ corresponds.
\begin{quote}
Now, when the pointer of the apparatus correctly measures the value of the observed quantity, we say that the measurement has happened. Therefore we can say that $M=1$ has the physical interpretation ``The measurement has happened,'' and $M= 0$ has the physical interpretation ``the measurement has not happened.'' (1037)
\end{quote}

Note, however, that the measurement of the observable $O$ (the measurement whose time we are concerned with) is to be described as an interaction between the system and apparatus, modeled by a time-dependent Hamiltonian.  If it remains isolated (i.e., without a measurement of $M$ being performed), Rovelli assumes that an ideal measurement interaction will take the uncorrelated state $\Psi(0)$ to the correlated, post-measurement state $\Psi(T)$ in way such that
\[  P(t) = \langle \Psi(t) | M \Psi(t) \rangle \] 
``will be a smooth function that goes monotonically from 0 to 1 in the time interval 0 to $T$'' (1037).\footnote{Now,  \cite{oppenheim2000does}  point out that there is a technical problem here since the system may be subject to the so-called backflow effect, in which case $P(t)$ can decrease with increasing $t$ and thus fails to be monotonic. There is also the further problem that no unitary evolution that can take a pure state to a mixture.  My concern, however, is conceptual rather than technical.}

To find out whether or not the measurement of $O$ has happened (as of time $t$), we must perform a measurement of $M$ on the combined system whose result indicates either a perfect correlation between system and apparatus, or no correlation at all.  Rovelli's proposal, then, involves two distinct types of measurement: there is the measurement of $O$ by the apparatus, described by unitary evolution of the joint system, and the measurement of the projector $M$ by the observer at a time of her choosing, described according to von Neumann's collapse process.

But if collapse (due to measurement of $O$) has already happened when $M$ is measured then $P(t)$ gives the wrong probabilities since it does not account for this fact.  And if it has not, then Rovelli's interpretation of $M$ as answering the question ``has a measurement of $O$ already occurred?'' cannot be sustained. That is, Rovelli's proposal faces the following dilemma:  

When measuring $M$ at time $t$ \emph{either} measurement of $O$ already occurred at some time $t'<t$, in which case $P(t)$ does not give the correct probabilities, \emph{or} measurement of $O$ occurs at time $t$, in which case the result $M=1$  cannot be interpreted as saying that a measurement has \emph{already} happened.

On neither of these alternatives does Rovelli's proposal do the job as advertised, and on the second we again reach the problematic conclusion that the experimenter may choose exactly when a measurement occurs.  This problem is avoided in taking the approach that I suggested above, which defines the time of occurrence as a conditional probability for occurrence within a time interval.  My way of thinking is also suggested by one of Rovelli's own formulations of the question we are trying to answer:
\begin{quote}
When is it precisely that a quantum event ``happens?'' Namely, when precisely can we replace the statement ``this may happen with probability $p$'' with the statement ``this has happened?''  (1032)
\end{quote}

His second formulation can be answered as follows: the statement ``this (event) has happened'' is true at times later than the occurrence of the event (as indicated by the use of the past tense).  Furthermore, the statement ``this event has happened'' becomes true not because the experimenter \emph{does} something to the system at some specific time, but simply because the event in question \emph{does occur} in the course of the experiment, and the time at which it does so is not under her control.  So instead of looking for an `experimental question' that we \emph{ask} at a specific time of the system, we should seek to investigate the question, ``at what time does the event occur?'' since it is after this time that we can truly say that the event has happened.   

What Rovelli's analysis of the situation here simply misses, however, is that the probabilistic answers given to questions about the time of occurrence for some event take it for granted \emph{that the event does occur} (at some unknown time), and so the statement ``this event will happen'' remains true until then.\footnote{
In the background here is a simple picture of temporal passage in which a future event becomes present and then past, and tensed statements about that event change their truth value in response.  (See \cite{prior1962changes} for the canonical statement of this view.) For example, if I light the fuse of a firework at time $t_b$ and say ``the firework will explode,'' that statement is true and is made true by the occurrence of the explosion at a later time $t_e>t_b$.   After the explosion, at times $t > t_e$, the statement ``the firework will explode'' is false while the statement ``the firework did explode'' is true, and is made true by the occurrence of the explosion at an earlier time.}  The reason that this alternative is hard to see is because the method of asking and answering questions \emph{at a time} is so ingrained in the von Neumann-Mackey formalism.  However, the idea that we should ask questions \emph{about} the time of an event (rather than at a time) arose naturally in giving a probabilistic analysis of atomic decay---a canonical example of a truly quantum process---and, in the previous section, I showed how to derive a probability distribution for the time of any given registration event arising from a quantum process.

\subsection{Oppenheim, Reznick and Unruh: Rinse and Repeat}

Oppenheim, Reznick and Unruh (\cite{oppenheim2000does}) accept Rovelli's interpretation of a measurement of $M$ as answering the question ``has the measurement of $O$ already occurred?''  However, they object that a single measurement of $M$ with result $M=1$ gives a poor determination of when the measurement of $O$ happened.  In order to isolate that time, they consider a family of repeated measurements of $M$.  The idea behind this is simple: to get a better read on this time, we record the results of many successive measurements of $M$ at times $\{t_1, t_2, t_3,\dots\}$.  Looking at a typical string of results of these measurements we will see something like $\{0,0,1,\dots\}$, which in this case indicates that the outcome occurred sometime between $t_2$ and $t_3$.

This, they claim, provides a more appropriate answer to the question ``when did the measurement occur?'' than Rovelli's single measurement of $M$.\footnote{They also point out that their analysis is completely general and applies to any projector.} To get the correct measurement statistics for successive measurements of $M$, however, requires taking into account the results of each prior  measurement. These joint probabilities require distinctive measurement statistics, arrived at by successive applications of L\"uders' Rule (\ref{luders_rule}).  

In particular, if the state is pure then $\rho = |\phi \rangle\langle \phi |$ for some unit vector $|\phi\rangle$ and so from (\ref{luders_rule}) we obtain
\begin{equation}
 \Pr ( E | F ) = \langle \phi' | E \phi' \rangle =  \frac{\langle \phi | F E F \phi \rangle }{\langle \phi | F \phi \rangle}
 \end{equation}
which leads to
\begin{equation}
 |\phi \rangle \rightarrow |\phi'\rangle = \frac{ F |\phi \rangle }{\sqrt{\langle \phi | F \phi \rangle}}.
\end{equation}

Now, for a measurement of $M$ at time $t$ there is an associated Heisenberg picture projector
\[  M_t = U_t^\dag M U_t, \]
where $U_t=e^{-iHt}$ is the one-parameter unitary group generated by the Hamiltonian $H$, a self-adjoint operator.  Furthermore, to any projector $P$ there corresponds an orthogonal projector, $P^0=\mathbb{I}-P$, associated with a measurement of eigenvalue 0 of $P$.\footnote{That is, the subspace onto which $\mathbb{I}-P$ projects is just the orthogonal complement of the range of $P$.}

Therefore for a pure state $|\Psi\rangle$ the joint probability for a sequence of successive measurements of $M$ at times $\{t_1,t_2, t_3\}$ with the results $\{0,0,1\}$ is given by
\begin{equation}
  \Pr(M=1\mbox{ at }t_3 | (M=0\mbox{ at }t_1)\; \& \; (M=0 \mbox{ at } t_2) )  = \langle \Psi'' | M_{t_3} \Psi''\rangle
\end{equation}
where $|\Psi''\rangle$ results from successive L\"uders state transitions induced by measurement of $M$ as follows:
\begin{equation}
 |\Psi\rangle \xrightarrow{M=0\mbox{\small{ at }}t_1} |\Psi'\rangle \xrightarrow{M=0\mbox{\small{ at }}t_2} |\Psi''\rangle \end{equation}
with
\begin{eqnarray}
|\Psi'\rangle =&   \frac{ (\mathbb{I} - M_{t_1}) | \Psi \rangle }{\sqrt{\langle \Psi | (\mathbb{I} - M_{t_1} )\Psi \rangle}} \\
|\Psi''\rangle =&  \frac{( \mathbb{I} - M_{t_2}) |\Psi' \rangle } {\sqrt{\langle \Psi' | ( \mathbb{I} - M_{t_2} )\Psi' \rangle} }    =   \frac{  (\mathbb{I} - M_{t_2}) (\mathbb{I} - M_{t_1})| \Psi \rangle }{\sqrt{\langle \Psi | (\mathbb{I} - M_{t_1}) (\mathbb{I} - M_{t_2}) (\mathbb{I} - M_{t_1}) \Psi \rangle}} .
 \end{eqnarray}

Oppenheim, Reznick and Unruh's proposal is just a special case of this general rule for calculating joint probabilities: according to their proposal, the probability that the time of the outcome of the experiment lies between two times is given by the application of L\"uders' Rule to a uniform sequence of instantaneous measurements of $M$ on the joint system (with original state $| \Psi\rangle$). That is, one chooses a time-resolution for the measurements and calculates the probability that the transition between $M=0$ and $M=1$ occurs within a given time interval.\footnote{One of the more counterintuitive aspects of this proposal is that successive measurements of $M$ need not agree---even after an outcome has ostensibly obtained---and thus a sequence of results such as $\{0,0,1,1,0\}$ is possible (in the sense that it is assigned non-zero probability). If a measurement of $M$ with result $M=1$ takes place at $t_0$ then the probability of measuring $M=1$ at some later time $t>t_0$ is given by 
\[ P(t) = \langle \Psi | M_{t_0} M_t M_{t_0} \Psi \rangle .\]
Assume for contradiction that the second measurement of $M$ is guaranteed to return $M=1$.  Then $P(t)=1$ for all $t>t_0$ and thus $M_{t_0}|\Psi\rangle $ is an eigenstate of $M_t$ with eigenvalue 1 for any $t>t_0$, that is
\[ M_t M_{t_0} |\Psi\rangle = M_{t_0}| \Psi\rangle .\]
Since $|\Psi\rangle$ was arbitrary, this is true for any $| \Psi\rangle \in \mathcal{H}$. In that case, it follows that $M_{t_0}\leq M_t$ (i.e., $M_{t_0}$ projects onto a subspace of $M_t$) and so they commute.  But, as Oppenheim, Reznick and Unruh point out (p.\ 133), in general two projections from the same family $M_t, M_{t'}$ with $t\neq t'$ do \emph{not} commute. (And, as they argue, exceptions to this claim will be rare.  For example, if the Hamiltonian is periodic then only if $|t-t'|$ is equal to (a multiple of) the period will $M_t, M_t'$ commute (since in that case $M_t = M_{t'}$).) Therefore, in general there is a non-zero probability that a measurement of $M_t$ will return $M=0$. QED. (Note that we can run the same argument using the Heisenberg picture family $M^0_t = U_t^\dag M^0 U_t = U_t^\dag(\mathbb{I} - M) U_t = \mathbb{I} - M_t$, or any such family.)}

However, the events to which one is really assigning probabilities are sets of outcomes of a series of instantaneous measurements of $M$ taking place at pre-determined times.  This is not a probability distribution for the time of an event.\footnote{Oppenheim, Reznick and Unruh critique Rovelli's proposal on the grounds that:
\begin{quote}
His scheme only answers the first question: ``has the measurement occurred already at a certain time?'', but does not answer the more difficult question ``when did the measurement occur?'' In other words, it does not provide a proper probability distribution for the time of an event. (108)
\end{quote}
I agree, but note that their proposal is subject to precisely this latter critique.}
Nonetheless, in taking this approach, which gives a  joint probability for the results of series of instantaneous measurements taking place at pre-defined times, they do avoid the assumption that the measurement process takes place at a time chosen by the experimenter. But this leads instead to the idea that the apparatus is repeatedly measuring $M$ of its own accord, which results in the (so-called) quantum Zeno effect.

\subsection{The Quantum Zeno Effect}\label{quant_zeno}

The foregoing derivation of probabilities for repeated measurement from the repeated application of  L\"uders' Rule is precisely the framework assumed by \cite{misra1977zeno} in their rigorous derivation of the quantum Zeno effect, which they present as resulting from their definition of ``continuous observation'' of a system over an interval of time.  According to them, continuous observation of some property of a system over an entire interval of time period $[0,\tau]$ can be modeled by considering the continuous limit of a uniform sequence of repeated measurements.

In particular, they assume that $n+1$ instantaneous measurements of some projector $P$ take place at regular intervals during the interval $[0,\tau]$  and that the probability that each of these measurements returns 0 is given by $n$ successive applications of L\"uders' Rule.  They prove (with all due rigor) that in taking the limit $n\rightarrow \infty$ this joint probability tends to zero.\footnote{They make two additional non-trivial assumptions.  First, they assume that the projection $P$ commutes with the one-parameter semi-group of time translations generated by the dynamics.  Second, they assume that the semi-group representing the evolution of the system under continuous observation (that results from taking the $n \rightarrow \infty$ limit) is continuous for $t\geq 0$, not just $t>0$.} They conclude:
\begin{quote}
We are thus led to the paradoxical conclusion that an unstable particle will not decay as long as it is kept under continuous observation as to whether it decays or not. \citep[p.\ 390]{misra1977zeno}
\end{quote}
Since unstable particles do seem to decay, the obvious inference to make is that such particles are not under continuous observation (in this sense). Misra and Sudarshan make several suggestions along these lines, one of which is that the sort of apparatus actually used to make observations in the lab (like a bubble chamber) is better modeled by the discrete repeated measurement process rather than its continuous limit.

This appears to be the opinion of Oppenheim, Reznick and Unruh, who extend this idea to suggest that there is ``always an inherent inaccuracy when measuring the time that the event (of measurement) occurred'' \cite[p.\ 177]{oppenheim2000does}. That is, the accuracy to which one can determine the time of measurement depends on the frequency with which one can measure $M$, and this frequency cannot be increased arbitrarily without running into the quantum Zeno effect (and thus ensuring that $M$ never comes to have value $M=1$).  However,  the conclusion that the quantum Zeno effect limits the accuracy to which one can determine the time of an event in this way depends crucially on the assumption that the experiment in question is best modeled by the repeated application of a projector to the system state.

Here I disagree: although \emph{some} experiments are suitably modeled in this way (see \cite{beige1996projection}), these are not experiments where the time of an outcome is allowed to vary. What the numerous purported demonstrations of the quantum Zeno effect really demonstrate is that this effect only comes into play under special experimental circumstances involving strong interactions between system and apparatus, often taking place at pre-determined times.  However, these circumstances do not arise when using registration devices such as photodetectors, Geiger counters, scintillators, bubble chambers, etc., and so a typical detector is poorly modeled by these means.

\cite{sudbery2002verdammte} argues for a similar conclusion by comparing the seminal demonstration of the quantum Zeno effect of \cite{itano1990quantum} with similar experiments aimed at displaying `quantum jumps' \citep{nagourney1986shelved}.  In the former case, one `measures' the presence of a system (a Be$^+$ ion) in its ground state through the application of a pulsed laser  (which induces fluorescence) and finds that as the frequency of pulses within some time period increases, the system is less likely to have made it out of the ground state during that period.  This is given the interpretation that the dynamics are `frozen' by the act of repeated measurement---the quantum Zeno effect.

In the quantum jump experiment of \citep{nagourney1986shelved}, however, the presence of the system in its ground state is detected similarly (through fluorescence) but the laser is not pulsed and couples less strongly.  In this case, the system displays the characteristically quantum behavior of `jumping' between the two energetically favored states at times that are fairly well-defined experimentally (although the distribution of these times, being essentially random, can only be predicted probabilistically).  Here, the continuous application of the laser has no Zeno-like effect.

Sudbery attributes the difference between these two cases to the fact that in the quantum jump experiment the laser is steady and so can be modeled by a time-independent interaction, whereas in the quantum Zeno experiment the pulsed laser requires a time-dependent interaction.  However, this cannot be the whole story since then we would expect to see no Zeno effect when the laser is left on permanently rather than pulsed, which is not what happens in the quantum Zeno experiment.\footnote{In fact,  \cite{itano1990quantum} describe how they did exactly this to prepare the system in the ground state.}  Instead, I want to endorse Ballentine's suggestion that, in the experiment of \cite{itano1990quantum},
\begin{quote}
 the quantum Zeno effect is not really a consequence of continuous observation, but rather of an excessively strong perturbation by an inappropriate apparatus'' \cite[p.\ 5166]{ballentine1991comment}
 \end{quote}
This point of view is confirmed by the recent experiment of \cite{patil2015measurement} which displays a clear separation between a ``Zeno regime'' of strong coupling and a weak coupling regime.

By design, then, a detector must be weakly coupled to the experimental system, and this coupling can be modeled by a time-independent interaction---crucially, without breaking the time translation symmetry of the system by applying repeated projections at pre-determined times. In that case, we may apply the method from Section \ref{cond_prob} to derive a conditional probability distribution for the time of occurrence from an effect that describes the event in question.\footnote{I note that \cite{ruschhaupt2009detector} provide a theoretical basis for modeling a detector as involving quantum jumps in a related way, and also make use of the operator normalization technique of \cite{brunetti2002time} to derive time of arrival POVMs. It seems likely that these POVMs could be interpreted as conditional probabilities (as suggested above), although \cite{ruschhaupt2009detector} do not interpret them as such.}  Here, while the probability of detection in unit time will vary with time, it varies as a result of unitary evolution of the system rather than an explicit time-dependence of the coupling.\footnote{To illustrate: if the interaction term of the Hamiltonian is position dependent then the strength of the coupling may vary with time (through the Schr\"odinger picture evolution of the joint state) but the Hamiltonian for the joint system will be time-independent.} 

This allows us to resolve the difficulties with predicting the time of an outcome along the lines suggested above: rather than thinking of a measurement as something one does to the system, we may think of an experiment as something that happens---a process that results in a special kind of quantum jump which leads to a registration event.\footnote{The focus of this paper is the time of this event, which arises from the weak coupling of two systems.  An alternative question concerns the correct description of the occurrence of many such events as a \emph{stochastic} process.  The use of a stochastic master equation to describe continuous weak measurements seems promising in this regard \cite[\S4]{jacobs2006}.}  This allows us to replace the idea that a measurement result arises from the application of an observable to the system at a pre-specified time with the idea that a registration event occurs as the result of a quantum jump of the detector at an essentially random (although probabilistically predictable) time.   In this way, the time translation symmetry of the system is broken by nature, not the experimenter.

\bibliography{Pashby_Zurich}

\end{document}